\documentclass[conference, draftcls, onecolumn]{IEEEtran}
\IEEEoverridecommandlockouts
\usepackage{cite}
\usepackage{amsmath,amssymb,amsfonts}
\usepackage{algorithm}
\usepackage[caption=false,font=footnotesize]{subfig}
\usepackage{algorithmic}
\usepackage{graphicx}
\usepackage{textcomp}
\usepackage{xcolor}
\def\BibTeX{{\rm B\kern-.05em{\sc i\kern-.025em b}\kern-.08em
    T\kern-.1667em\lower.7ex\hbox{E}\kern-.125emX}}
    \usepackage[letterpaper, left=0.65in, right=0.65in, bottom=1in, top=0.71in]{geometry}
\begin{document}

\title{Predictive Relay Selection: A Cooperative Diversity Scheme Using Deep Learning\\
{\footnotesize}
}

\author{\IEEEauthorblockN{Wei Jiang}
\IEEEauthorblockA{\textit{German Research Center for Artificial Intelligence (DFKI)}\\
Kaiserslautern, Germany \\
https://orcid.org/0000-0002-3719-3710}
\and
\IEEEauthorblockN{Hans Dieter Schotten}
\IEEEauthorblockA{ \textit{University of Kaiserslautern}\\
Kaiserslautern, Germany \\
https://orcid.org/0000-0001-5005-3635}
}

\maketitle

\begin{abstract}
In this paper, we propose a novel cooperative multi-relay transmission scheme for mobile terminals to exploit spatial diversity. By improving the timeliness of measured channel state information (CSI) through deep learning (DL)-based channel prediction, the proposed scheme remarkably lowers the probability of wrong relay selection arising from outdated CSI in fast time-varying channels. It inherits the simplicity of opportunistic relaying by selecting a single relay, avoiding the complexity of multi-relay coordination and synchronization. Numerical results reveal that it can achieve full diversity gain in slow-fading channels and substantially outperforms the existing schemes in fast-fading wireless environments. Moreover, the computational complexity brought by the DL predictor is negligible compared to off-the-shelf computing hardware.
\end{abstract}


\section{Introduction}
Cooperative diversity \cite{Ref_WJ_jiang2017device} is an effective technique for mobile terminals without an antenna array to cultivate spatial diversity that is typically achieved by co-located multi-antenna systems.  A main challenge of cooperative diversity is the inherent asynchronization among spatially-distributed antennas (relays). Multiple timing offset  and multiple carrier frequency offset \cite{Ref_SYN01} among simultaneously-transmitting relays make \emph{multi-relay} transmission such as distributed beam-forming and distributed space-time coding \cite{Ref_DSTC}  too complicated for practical systems. In contrast, a \emph{single-relay} approach called opportunistic relay selection (ORS) or opportunistic relaying  \cite{Ref_OPR} achieves full diversity gain while the complexity of multi-relay synchronization and coordination is avoided.

However, ORS is applicable only in slow-fading wireless environments since channel state information (CSI) used to select the best relay may be outdated quickly in fast-fading channels. Using a wrongly-selected relay substantially deteriorates the performance of ORS, as widely verified in the literature such as \cite{ Ref_myVTC2014, Ref_Vicario, Ref_WJ_jiang2014opportunistic}. With the proliferation of high-mobility applications (such as vehicle-to-X, high-speed train, and unmanned aerial vehicle) and the utilization of higher frequency bands (e.g., millimeter wave and Terahertz) in 5G and beyond systems, the problem of outdated/aged CSI becomes more challenging. A cooperative method called generalized selection combining \cite{Ref_Xiao} shows robustness under aged channel but it suffers from a substantial loss of spectral efficiency. The authors of \cite{Ref_Li_RelaySelection} proposed a method utilizing the knowledge of channel statistics, getting only a marginal gain, whereas the complexity obviously grows. By far, opportunistic space-time coding (OSTC) proposed by the author of this paper in \cite{Ref_MineTRANS, Ref_Jiang_ICC, Ref_WJ_jiang2014achieving} is the best method in fast fading channel from the perspective of diversity-multiplexing trade-off. But its performance gap to perfect selection using the perfect knowledge of channel is still large, motivating our follow-up works presented here.

In this paper, therefore, we propose a novel cooperative method coined predictive relay selection (PRS) for mobile terminals to exploit the gain of spatial diversity. The probability of wrong relay selection due to outdated CSI is remarkably reduced by improving the timeliness of CSI through fading channel prediction \cite{Ref_myIEEEAccess, Ref_myOJCOMS, Ref_JiangVTC2, Ref_JiangFDpredict, Ref_WJ_jiang2020neuralnetwork, Ref_WJ_jiang2018multi, Ref_WJ_jiang2020recurrent, Ref_WJ_jiang2020long, Ref_WJ_jiang2020deeplearning}. A deep recurrent neural network is deliberately built to provide high-accurate CSI predictions. The proposed scheme inherits the simplicity of ORS by selecting a single opportunistic relay to avoid the complexity of multi-relay coordination and synchronization. Simulation results reveal that it can achieve full diversity order in slow-fading channels and substantially outperforms the existing schemes in fast-fading wireless environments. Moreover, the computational complexity brought by the deep learning (DL)-based predictor is analyzed and compared with commercial off-the-shelf (COTS) computing hardware.
The rest of this paper is organized as follows: Section II introduces the system model. Section III and IV present the proposed selection scheme and the principle of channel predictor, respectively. Complexity analysis and numerical results are given in Section V and VI. Finally, Section VII concludes this paper.
\section{System Model}
Following the working assumption applied for most of prior research works \cite{Ref_DSTC, Ref_Jiang_ICC, Ref_SYN01,  Ref_OPR, Ref_Vicario, Ref_Xiao,  Ref_Li_RelaySelection, Ref_MineTRANS, Ref_WJ_jiang2014achieving, Ref_WJ_jiang2014opportunistic, Ref_myVTC2014}, we consider a dual-hop decode-and-forward (DF) cooperative network where a single source node $s$ communicates with a single destination node $d$ with the aid of $K$ relays. Each node is equipped with a single antenna that is used for both signal transmission and reception over a narrow-band channel.  The received signal in an arbitrary link $A{\rightarrow}B$ is modeled as $y_{B} = h_{A,B}x_{A} + z_{B}$, where $x_A\in \mathcal{C}$ is the transmitted symbol from node $A$ with average power $P_A=\mathbb{E}[|x_A|^2]$,  $z_{B}$ stands for additive white Gaussian noise with zero-mean and variance $\sigma^2_n$, i.e., $z {\sim} \mathcal{CN}(0,\sigma^2_n)$, and $h_{A,B}$ represents the fading coefficient of the channel from $A$ to $B$, which is a zero-mean circularly-symmetric complex Gaussian random variable  $h {\sim} \mathcal{CN}(0, \sigma_h^2)$ under the assumption of Rayleigh fading. The instantaneous signal-to-noise ratio (SNR)  is denoted by $\gamma_{A,B}{=}|h_{A,B}|^2 P_A /\sigma_n^2$ and the average SNR $\bar{\gamma}_{A,B}{=}\mathbb{E}[\gamma_{A,B}]{=} P_A\sigma_h^2/\sigma_n^2$. 

In a practical system, there exists a delay between the time of relay selection and the instant of using the selected relay to transmit signals. The actual CSI $h$ may differ from its outdated version $\hat{h}$ that is applied for selecting relays. To quantify the quality of CSI, the correlation coefficient between $h$ and $\hat{h}$ is introduced, i.e., $\rho_o=\frac{\mathbb{E}[h\hat{h}^*]}{\sqrt{\mathbb{E}[|h|^2] \mathbb{E}[|\hat{h}|^2]}}$.
With the classical Doppler spectrum of the Jakes model, it takes the value
\begin{equation}
\label{Eqn_Jakes}
\rho_o=J_0(2\pi f_d \tau),
\end{equation}
where $f_d$ is the maximal Doppler frequency, $\tau$ stands for the delay between the outdated and actual CSI, and $J_0(\cdot)$ denotes the $zeroth$ order  Bessel  function  of the first kind.
\begin{figure}[!bpth]
\centering
\includegraphics[width=0.5\textwidth]{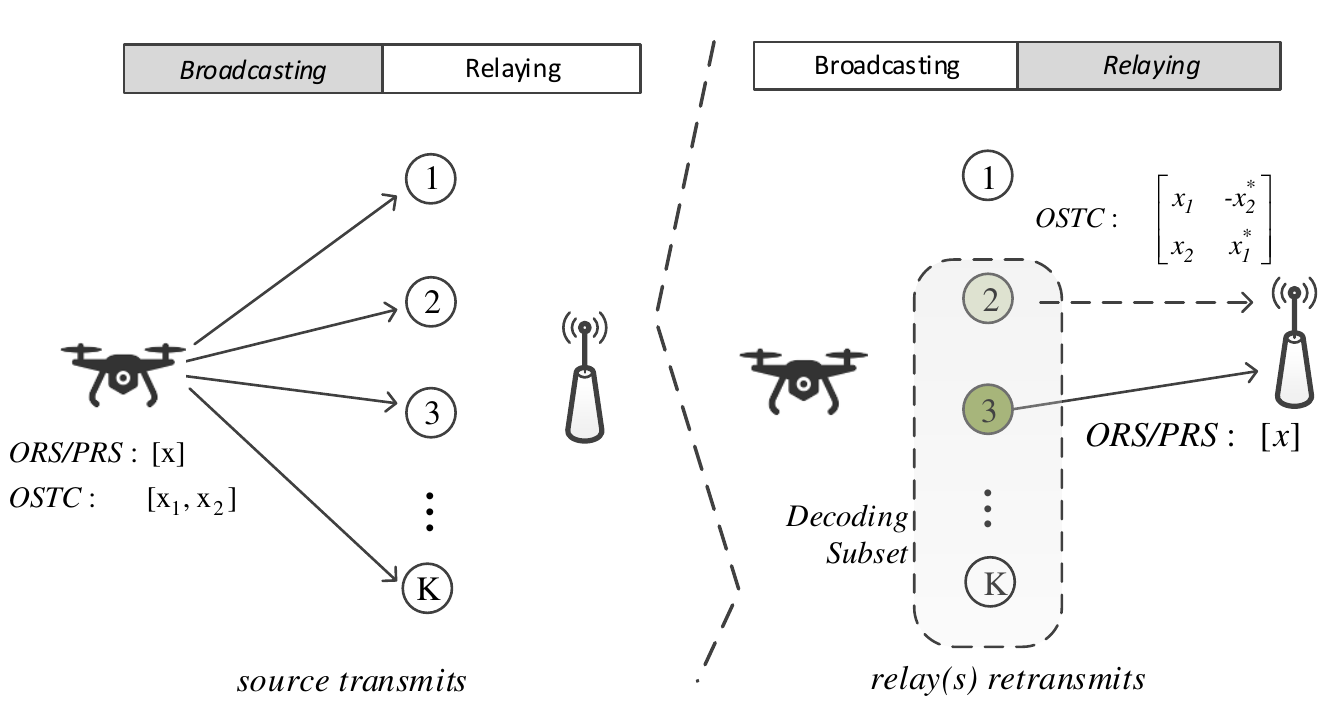}
\caption{Schematic diagram of a cooperative network using different DF relaying strategies: ORS, PRS, and OSTC. }
\label{Figure_RelaySele}
\end{figure}

Due to severe signal attenuation, a single-antenna relay should operate in half-duplex transmission mode to prevent from harmful self-interference between the transmitter and receiver. Therefore, its signal transmission is organized in two phases:  the source broadcasts a signal in the source-to-relay (denoted by $\mathbb{SR}$ hereinafter)  link, and then the relays retransmit this signal in the relay-to-destination ($\mathbb{RD}$) link.
In the first phase, as shown in \figurename \ref{Figure_RelaySele}, the source (e.g., the drone in the figure) sends a symbol $x$ and those relays who can correctly decode $x$  form a \emph{decoding subset} ($\mathcal{DS}$) of the $\mathbb{SR}$ link
\begin{equation}
 \label{Eqn_DS} 
\mathcal{DS}  \triangleq \left \{ k \left|  \log_2(1+\gamma_{s,k}) \geqslant 2R \right.  \right\},  
\end{equation}
where $R$ is an end-to-end ($\mathbb{EE}$) target rate for the dual-hop relaying. Note that the required data rate for either hop is doubled to $2R$ due to the adoption of half-duplex  transmission.
The best relay (denoted by $\dot{k}$) in ORS  is opportunistically selected from $\mathcal{DS}$ in terms of $\dot{k}=\arg \max_{k\in \mathcal{DS}} {\hat{\gamma}_{k,d}}$, where $\hat{\gamma}_{k,d}$ is the SNR of the $\mathbb{RD}$ link \emph{at the instant of relay selection}, which is an outdated version of the actual SNR $\gamma_{k,d}$ during signal transmission. In contrast, the proposed PRS scheme replaces the outdated CSI with the predicted CSI $\check{h}$, and determines $\dot{k}$ in terms of $\dot{k}=\arg \max_{k\in \mathcal{DS}} {\check{\gamma}_{k,d}}$, where $\check{\gamma}_{k,d}=|\check{h}_{k,d}|^2P_k/\sigma_n^2$. In addition to the best relay, the OSTC scheme \cite{Ref_MineTRANS} needs another relay with the second strongest SNR, i.e., $\ddot{k}=\arg \max_{k\in \mathcal{DS}-\{\dot{k}\}} {\hat{\gamma}_{k,d}}$.
In the first phase, the source broadcasts a pair of symbols $(x_1,x_2)$ to all relays over two consecutive symbol periods. The regenerated signals are encoded by means of the Alamouti scheme \cite{Ref_Jiang_ICC},  which is the unique space-time code achieving both full rate and full diversity, at the pair of selected relays. In the second phase, a relay transmits $(x_1,-x_2^*)$ while another transmits $(x_2,x_1^*)$ simultaneously at the same frequency.

\section{Predictive Relay Selection}
\begin{figure}[!h]
\centering
\includegraphics[width=0.5\textwidth]{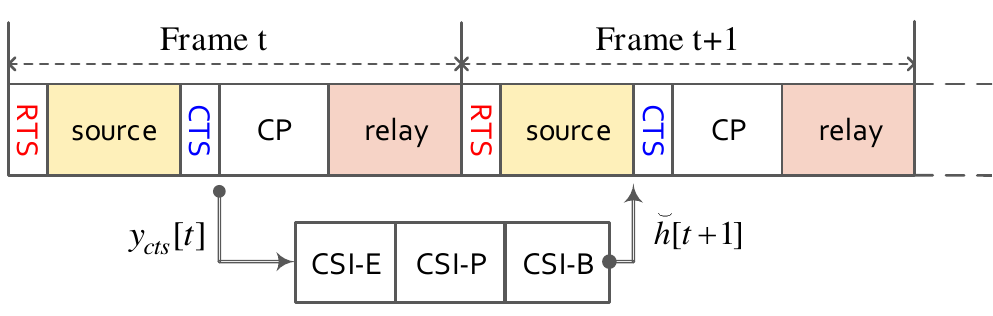}
\caption{Frame structure of PRS. \emph{CSI-E: CSI Estimation, CSI-P: CSI Prediction, CSI-B: CSI Buffering, CP: Contention Period}. }
\label{Figure_Implementation}
\end{figure}

Taking advantage of new degree of freedom opened by channel prediction, we propose the PRS scheme that can also achieve high performance in fast time-varying channels. The prediction horizon relaxes the tight requirement of time procedure and therefore provides the flexibility to design an advanced relaying strategy. As depicted in $\mathrm{Algorithm}~2$, the implementation for PRS is detailed as follows:
\begin{enumerate}[\IEEEsetlabelwidth{5)}]
\item
At frame $t$, as illustrated in \figurename \ref{Figure_Implementation}, the source broadcasts a packet containing a pilot called Ready-To-Send (RTS) \cite{Ref_OPR} and data payload. The CSI $h_{s,k}[t]$ is acquired at relay $k$ by estimating RTS and is used for detecting data symbols. Those relays that correctly decode the source's signal comprise a $\mathcal{DS}$.
\item
Clear-To-Send (CTS) is sent from the destination, so that  relay $k$ can estimate $h_{d,k}[t]$ and then $h_{k,d}[t]$ is known  due to channel reciprocity. It feeds $h_{k,d}[t]$ into its embedded channel predictor to generate $\check{h}_{k,d}[t+1]$, and buffers it for its usage at the upcoming frame $t+1$.
\item
Meanwhile, relay $k$ belonging to the $\mathcal{DS}$ fetches $\check{h}_{k,d}[t]$ that is buffered at the previous frame $t-1$. This operation starts once the CTS arrives, in parallel with Step $2$.
\item
Then, a timer with a duration $T_t$ proportional (denoted by $\propto$) to $1/|\check{h}_{k,d}[t]|$ is started at relay $k$. 
\item
The timer on the relay with the largest channel gain expires first, and then it sends a short packet to announce.
\item
Once received the best relay's notification, other relays terminate their timers and keep silent. The selected relay forwards the signal until the end of this frame.
\end{enumerate}

\begin{algorithm}
\label{Alg_1}
\caption{Predictive Relay Selection}
\begin{algorithmic}
\FOR{$t=1,2,...$}
\STATE $s$ sends RTS; $s$ sends data payload $\boldsymbol{x}[t]$
\FOR{$k=1,...,K$}
\STATE estimate $h_{s,k}[t]$;  $\hat{\boldsymbol{x}}[t]=f(\boldsymbol{y}_{s,k}[t],h_{s,k}[t])$
\IF {$\hat{\boldsymbol{x}}[t]$ is error-free}
\STATE fetch $\check{h}_{k,d}[t]$; start a timer $T_t\propto\frac{1}{|\check{h}_{k,d}[t]|}$ 
\ENDIF
\ENDFOR
\STATE $d$ sends CTS
\STATE $\dot{k}=\arg \max_{k\in \mathcal{DS}}\left(|{\check{h}_{k,d}[t]}|\right)$ notifies its presence
\STATE $\dot{k}$ transmits $\hat{\boldsymbol{x}}[t]$
\FOR{$k=1,...,K$}
\STATE estimate $h_{k,d}[t]$; predict $\check{h}_{k,d}[t+1]$
\STATE write $\check{h}_{k,d}[t+1]$ into Buffer
\ENDFOR
\ENDFOR
\end{algorithmic}
\end{algorithm}

\section{DL-based Channel Prediction}
This section first introduces the principle of deep recurrent networks including simple recurrent neural network (RNN) \cite{Ref_JiangFDpredict}, Long Short-Term Memory (LSTM) \cite{Ref_LSTM}, and Gated Recurrent Unit (GRU) \cite{Ref_GRU}, followed by the explanation of applying a recurrent network to build a channel predictor.

\subsection{Deep Recurrent Networks}

Unlike  uni-direction information flow in feed-forward neural networks,  RNN has recurrent self-connections, which are applied to memorize historical states, exhibiting great potential in time-series prediction. The activation of the previous time step is fed back as part of the input for the current step. In a simple RNN, its $l^{th}$ recurrent layer is generally modeled as
\begin{equation}
\label{Eqn_RNN_hiddenlayer}
\mathbf{d}_t^{(l+1)}=\mathcal{R}^{(l)}(\mathbf{d}_t^{(l)})=\delta_h \left( \mathbf{W}^{(l)} \mathbf{d}_t^{(l
)} + \mathbf{U}^{(l)} \mathbf{d}_{t-1}^{(l+1)} +\mathbf{b}^{(l)} \right),
\end{equation}
where $\mathbf{W}^{(l)}$ and $\mathbf{U}^{(l)}$ are weight matrices of the $l^{th}$ layer, $\mathbf{b}^{(l)}$ is a bias vector,  $\mathbf{d}_t^{(l)}$ and $\mathbf{d}_{t}^{(l+1)}$ represent the input and output for layer $l$ at time $t$, respectively, $\mathbf{d}_{t-1}^{(l+1)}$ is the feedback from the previous step, $\mathcal{R}^{(l)}(\cdot)$ stands for the relation function for the input and output of the $l^{th}$ RNN hidden layer, and the activation function  often selects  the \emph{hyperbolic tangent} denoted by $\mathrm{tanh}$, which is $\delta_h(x) =(e^{2x}-1)/(e^{2x}+1)$.
\begin{figure}[!thpb]
\centering
\includegraphics[width=0.41\textwidth]{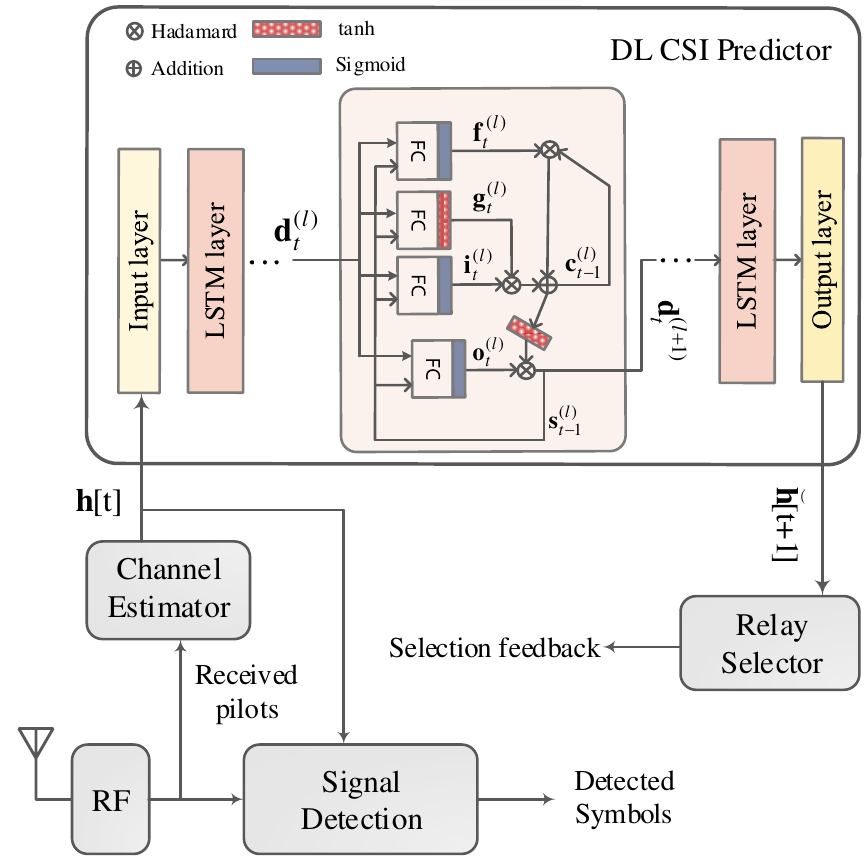}
\caption{Block diagram of the receiver of a relay in PRS. The DL-based predictor consists of an input layer, $L$ LSTM hidden layers, and an output layer, where the $l^{th}$ hidden layer is opened to illustrate the internal structure of an LSTM memory block. }
\label{Figure_DLpredictor}
\end{figure}
Using typical stochastic gradient descent method to train a recurrent network, the back-propagated error signals tend to zero that implies a prohibitively-long convergence time. To tackle this gradient-vanishing problem, Hochreiter and Schmidhuber proposed Long Short-Term Memory in their pioneer work of \cite{Ref_LSTM}, which introduced  \textit{cell} and \textit{gate} into the RNN structure. A typical LSTM cell has three gates: an \emph{input} gate controlling the extent of new information flows into the cell, a \emph{forget} gate  to filter out useless memory, and an \emph{output} gate that controls the extent to which the memory is applied to generate the activation.
The upper part of \figurename \ref{Figure_DLpredictor} shows the graphical depiction of a deep LSTM network consisting of an input layer, $L$ hidden layers, and an output layer.  Let's use the $l^{th}$ hidden layer as an example to shed light on how an activation signal goes through the network.  There are two hidden states - the short-term state $\mathbf{s}_{t-1}^{(l)}$ and the long-term state $\mathbf{c}_{t-1}^{(l)}$.  The input $\mathbf{d}_t^{(l)}$ and $\mathbf{s}_{t-1}^{(l)}$ jointly activate four fully connected (FC) layers, generating the activation vectors for the gates, i.e.,
\begin{equation} \left\{ \begin{aligned}
\label{Eqn_No1}
\mathbf{i}_t^{(l)} & = \delta_g \left( \mathbf{W}^{(l)}_{i}\mathbf{d}_t^{(l)} + \mathbf{U}_{i}^{(l)} \mathbf{s}_{t-1}^{(l)} +\mathbf{b}_i^{(l)} \right)\\
\mathbf{o}_t^{(l)}  &= \delta_g \left( \mathbf{W}_{o}^{(l)} \mathbf{d}_t^{(l)}  + \mathbf{U}_{o}^{(l)} \mathbf{s}_{t-1}^{(l)} +\mathbf{b}_o^{(l)} \right)\\
\mathbf{f}_t^{(l)} & = \delta_g \left( \mathbf{W}_{f}^{(l)} \mathbf{d}_t^{(l)} + \mathbf{U}_{f}^{(l)} \mathbf{s}_{t-1}^{(l)} +\mathbf{b}_f^{(l)} \right)
\end{aligned}, \right.
\end{equation}
where $\mathbf{W}$ and $\mathbf{U}$ are weight matrices for the FC layers, $\mathbf{b}$ represents bias, the subscripts  $i$, $o$, and $f$ associate with the input, output, and forget gate, respectively, and $\delta_g$ stands for the logistic \emph{Sigmoid} function $\delta_g(x) = 1/(1+e^{-x})$.
The  current  long-term state  $\mathbf{c}_t^{(l)}$ is obtained by first throwing away outdated memory at the forget gate and then adding new information selected by the input gate, i.e., $
\mathbf{c}_t^{(l)}  =  \mathbf{f}_t^{(l)} \otimes \mathbf{c}_{t-1}^{(l)} +  \mathbf{i}_t^{(l)} \otimes  \mathbf{g}_t^{(l)}$,
where the operator $\otimes$ denotes the Hadamard product (element-wise multiplication) and $\mathbf{g}_t^{(l)}=\delta_h ( \mathbf{W}_{g}^{(l)} \mathbf{d}_t^{(l)} + \mathbf{U}_{g}^{(l)} \mathbf{s}_{t-1}^{(l)} +\mathbf{b}_g^{(l)} )$.
The output of this hidden layer is computed by
\begin{equation}
\label{Eqn_LSTMinputoutput}
 \mathbf{d}_t^{(l+1)}  = \mathcal{L}^{(l)} \left(\mathbf{d}_t^{(l)}\right)= \mathbf{o}_t^{(l)} \otimes \delta_h \left( \mathbf{c}_{t}^{(l)} \right),
\end{equation}
where $\mathcal{L}^{(l)}(\cdot)$ represents the input-output function for the $l^{th}$ LSTM layer. 

Despite of its short history, LSTM has achieved a great success and been commercially applied in many AI products such as Apple Siri and Google Translate. Since  its emergence,  the research community published a number of its variants, among which GRU proposed by Cho \textit{et al.} in \cite{Ref_GRU} draws lots of attention.  It's a simplified version with fewer parameters, but it exhibits even better performance over LSTM on certain smaller  and less frequent datasets. To simplify the structure, a GRU memory cell has only a single hidden state, and the number of gates is reduced to two: the \emph{update} and \emph{reset} gate. The activation vector for the update gate is computed by $\mathbf{z}_t^{(l)}  = \sigma_g ( \mathbf{W}_{z}^{(l)} \mathbf{d}_t^{(l)} + \mathbf{U}_{z}^{(l)} \mathbf{s}_{t-1}^{(l)} +\mathbf{b}_z^{(l)} )$, which decides the extend to which the memory content from the previous state will remain in the current state.
The reset gate controls whether the previous state is ignored, and when it tends to $0$, the hidden state is reset with the current input. It is given by $\mathbf{r}_t^{(l)}  = \sigma_g ( \mathbf{W}_{r}^{(l)}\mathbf{d}_t^{(l)} + \mathbf{U}_{r}^{(l)} \mathbf{s}_{t-1}^{(l)} +\mathbf{b}_r^{(l)} )$. Likewise, the previous hidden state $ \mathbf{s}_{t-1}^{(l)}$ goes through the cell, drops outdated memory, and inserts some now content, generating the current hidden state, that is
\begin{align}\label{Eqn_No2}
\mathbf{s}_t^{(l)} & =   (1- \mathbf{z}_t^{(l)}) \otimes \mathbf{s}_{t-1}^{(l)}\\ \nonumber &+  \mathbf{z}_t^{(l)} \otimes \sigma_h \left( \mathbf{W}_{s}^{(l)}\mathbf{d}_t^{(l)} + \mathbf{U}_{s}^{(l)}( \mathbf{r}_t^{(l)} \otimes \mathbf{s}_{t-1}^{(l)}) +\mathbf{b}_s^{(l)} \right).
\end{align}
The hidden state is also equal to its output of this hidden layer, i.e., $\mathbf{d}_t^{(l+1)}  = \mathcal{G}^{(l)}(\mathbf{d}_t^{(l)})=\mathbf{s}_t^{(l)}$, where $\mathcal{G}^{(l)}(\cdot)$ denotes the input-output function.

\subsection{DL-based Channel Predictor}

To shed light on the principle of a DL-based predictor, as shown in \figurename \ref{Figure_DLpredictor}, the chain of signal reception at the receiver is demonstrated. The predictor is inserted at the end of a channel estimator and generates predicted CSI to replace outdated CSI as the input for a relay selector. It is transparent and therefore an ORS system can be smoothly upgraded to a PRS system without any other modifications.  In such a \emph{distributed-selection} method,  each relay requires to process only local CSI $h_{k,d}[t]$.  As we know, a complex-valued fading coefficient can be expressed in polar form as $h_{k,d}[t]=a_{k,d}[t]e^{j\theta_{k,d}[t]}$, where $a_{k,d}[t]$ and $\theta_{k,d}[t]$ denote the magnitude and phase, respectively. Because the selection relies on the value of SNR, only the knowledge of magnitude $a_{k,d}[t]$ is enough, rather than complex-valued $h_{k,d}[t]$, which in turn can simplify the implementation of the channel predictor by employing  a neural network with real-valued weights and biases. Feeding  $a_{k,d}[t] $ into the input feed-forward layer obtains one-dimensional output $\mathbf{d}^{(1)}_t=d_t^{(1)}=\delta_h ( w^{(i)} a_{k,d}[t] +b^{(i)} )$, where $w^{(i)}$ and  $b^{(i)}$ denote the weight and bias of the input layer. The activation of the $1^{st}$ hidden layer is exactly $\mathbf{d}^{(1)}_t$, then $\mathbf{d}_t^{(2)}=\mathcal{L}^{(1)}(\mathbf{d}_t^{(1)})$ is generated and forwarded to the $2^{nd}$ hidden layer, where $\mathcal{L}^{(1)}\left(\cdot\right)$ is defined in (\ref{Eqn_LSTMinputoutput}). The activation goes through the network until the output layer gets the predicted CSI, which is computed by $\check{a}_{k,d}[t{+}1]=\delta_h ( \mathbf{W}^{(o)} \mathbf{d}_t^{(L)} +b^{(o)} )$, where $\mathbf{W}^{(o)}$ and  $b^{(o)}$ denote the weight matrix and bias of the output layer, and the activation of the last hidden layer equals to $\mathbf{d}_t^{(L)}=\mathcal{L}^{(L)}( \ldots \mathcal{L}^{(2)}(\mathcal{L}^{(1)}(\mathbf{d}_t^{(1)})))$. The building of a deep recurrent network is flexible, for example, we can apply a hybrid network consisting of  RNN, GRU, and LSTM layers, like $\mathbf{d}_t^{(L)}=\mathcal{G}^{(L)} ( \ldots \mathcal{L}^{(2)} (\mathcal{R}^{(1)}(\mathbf{d}_t^{(1)})))$.

\section{Computational Complexity}
In the context of cooperative diversity, the computational complexity mainly arises from multi-relay coordination and synchronization \cite{Ref_SYN01}. The simplicity of ORS is achieved thanks to single-relay transmission that substantially lowers the amount of signalling overhead among multiple relays. A direct comparison of different schemes is not easy and does not provide real insight. That is why most of the works in this field \cite{Ref_DSTC, Ref_Jiang_ICC, Ref_SYN01,  Ref_OPR, Ref_Vicario, Ref_Xiao,  Ref_Li_RelaySelection, Ref_MineTRANS, Ref_WJ_jiang2014achieving, Ref_WJ_jiang2014opportunistic, Ref_myVTC2014} did not provide a quantitative analysis. On the other hand, the complexity of the proposed scheme comes mainly from the DL-based predictor, which is always a concern for the application of deep learning.  From a practical perspective, it is more meaningful to make clear its demand on computing resources  in comparison with the capability of COTS hardware. Hence, let's focus on assessing the complexity of the predictors in terms of floating-point operations per second (FLOPS).

A deep recurrent network can be quantitatively modelled as follows: an input layer with $N_i$  neurons, an output layer with $N_o$ neurons, and $L$ hidden layers, which has $N_h^{l}$ neurons at layer $l=1,\ldots,L$. To begin with the input layer, it computes $\delta_h ( \mathbf{W}^{(i)} \mathbf{d} +\mathbf{b}^{(i)} )$, where the matrix multiplication generates $N_iN_h^1$ floating-point multiplicative operations and  $(N_i-1)N_h^1$ additive operations, and the addition of the bias vector consumes $N_h^1$ operations, amounting to  a total of $2N_iN_h^1$. Note that the amount of computation raised by the activation function is negligible compared to the matrix multiplication, which is usually ignored in the calculation of complexity for deep learning. Likewise, it is easy to know that the output layer corresponds to $2N_h^LN_o$. For an RNN hidden layer as given in (\ref{Eqn_RNN_hiddenlayer}), the number of operations equals to $O^{l}=(2N_h^{l-1}-1)N_h^l+(2N_h^{l}-1)N_h^l+N_h^{l}$,
where the first term corresponds to the calculation of $ \mathbf{W}^{(l)} \mathbf{d}_t^{(l
)}  $, the second is for $ \mathbf{U}^{(l)} \mathbf{d}_{t-1}^{(l+1)}$, and the third is due to the addition of the bias.  For simplicity, $O^{l}$ can be approximated to $2N_h^{l-1}N_h^l+2(N_h^l)^2$. Then, the overall complexity for a simple RNN is given by
\begin{equation}
    \label{Eqn_complexity_RNN}
O_{rnn}\approx2\left[ N_iN_h^1+N_h^LN_o +\sum_{l=1}^L\left( N_h^{l-1}N_h^l+\left(N_h^l\right)^2 \right)  \right],
\end{equation}
where we apply $N_h^{0}=N_i$ for a simpler expression.
As derived from (\ref{Eqn_No1})-(\ref{Eqn_LSTMinputoutput}), the number of operations for the matrix multiplication on an LSTM layer is $4$ times that of an RNN layer, i.e., $4O^l$. The computation for the gate control, which has totally $7N_h^l-3$ operations, can be neglected. Therefore, the complexity of an LSTM network is approximated by
\begin{equation}
\label{Eqn_complexity_deepLSTM}
O_{lstm} \approx 2\left[N_iN_h^1+N_h^LN_o + \sum_{l=1}^{L} 4\left( N_h^{l-1}N_h^l+\left(N_h^l\right)^2 \right)\right].
\end{equation}
 Similarly, we can derive the expression for GRU, i.e.,
 \begin{equation}
 \label{Eqn_complexity_GRU}
O_{gru} \approx 2\left[N_iN_h^1+N_h^LN_o + \sum_{l=1}^{L} 3\left ( N_h^{l-1}N_h^l+\left(N_h^l\right)^2 \right)\right]
\end{equation}
Note that the above expressions are the complexity per prediction step, we need to multiply (\ref{Eqn_complexity_RNN})-(\ref{Eqn_complexity_GRU}) with the frequency of prediction denoted by $f_p$, i.e., the number of steps performed per second,  to figure out FLOPS. 

Given the concrete values of these parameters, the complexity of the predictor is quantified to compare with the capacity of COTS computing hardware. Suppose the applied deep neural network has two LSTM hidden layers with $N_h^1=N_h^2=25$ neurons\footnote{The selection of such hyper-parameters will be justified in the next section.}. The input for the predictor at the $k^{th}$ relay is $a_{k,d}[t]$, corresponding to $N_i=N_o=1$. It amounts to $O_{lstm}=15,300$ floating-point operations \emph{per prediction} in terms of (\ref{Eqn_complexity_deepLSTM}). The interval of prediction step is assumed to be $1\mathrm{ms}$,  the frequency of prediction equals to $f_p=1,000$, resulting in $15.3\mathrm{MFLOPS}$. In comparison with off-the-shelf Digital Signal Processors (DSPs), e.g., TI C6678, which provides a computation capacity of up to $179\mathrm{GFLOPS}$, the required computing resource occupies less than $0.01\%$ of a single DSP chip. Taking into account its back-compatibility to legacy hardware, we further check low-end DSPs. Given TI C6748 that has  computation power of $2.7\mathrm{GFLOPS}$ as an example, the resource required by the predictor is around $0.6\%$.  In a nutshell, the complexity of the DL-based channel predictor applied for PRS is quite affordable, if not negligible.

\section{Simulation results}

\begin{figure*}[!t]
\centerline{
\subfloat[]{
\includegraphics[width=0.325\textwidth]{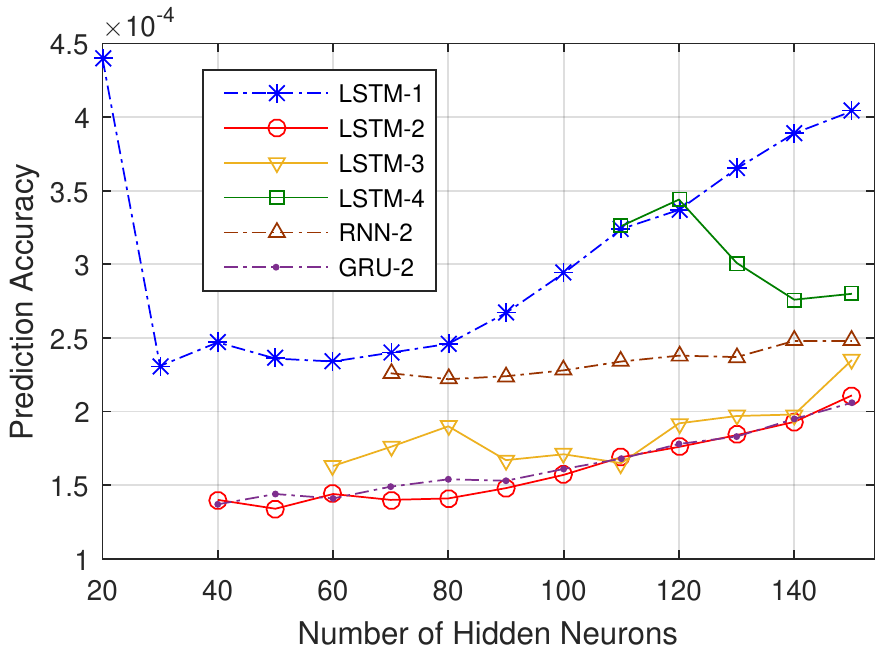}
\label{Fig_MSE}
}
\hspace{0mm}
\subfloat[]{
\includegraphics[width=0.31\textwidth]{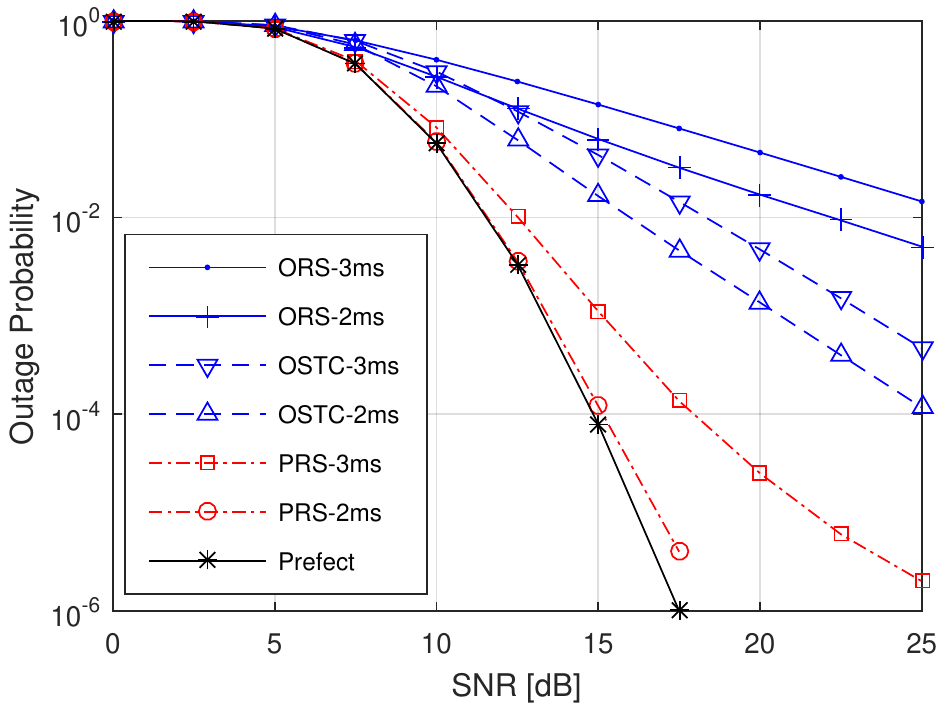}
\label{Fig_DF_outage}
}
\hspace{0mm}
\subfloat[]{
\includegraphics[width=0.31\textwidth]{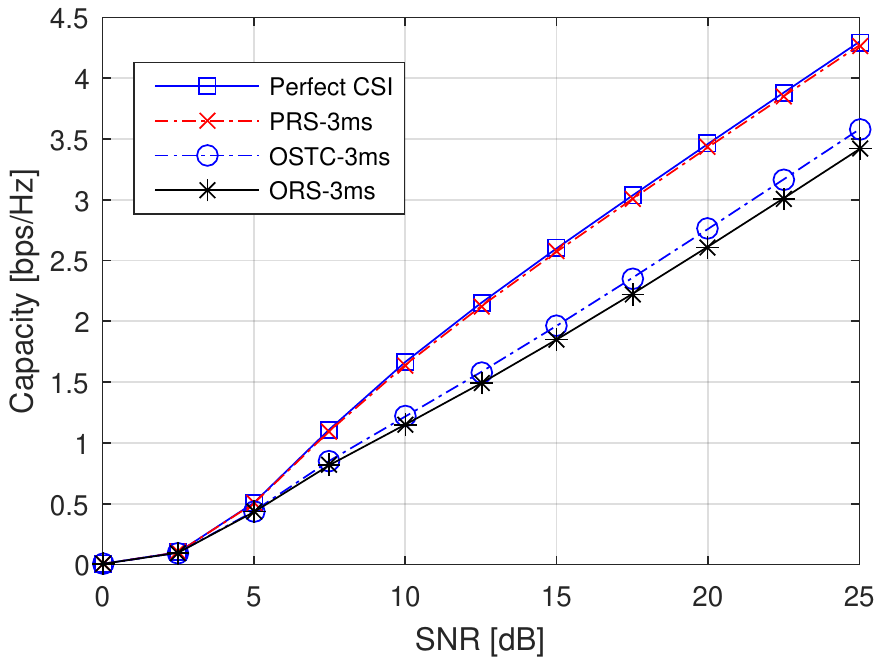}
\label{Fig_capacity}
}}
\hspace{15mm}
\caption{(a) Prediction accuracy in terms of the number of hidden neurons. (b) Comparison of outage probability for ORS, OSTC, and PRS in a  cooperative network with $K{=}8$ relays; (c) Comparison of channel capacity for ORS, OSTC, and PRS in a cooperative network with $K{=}8$ relays.}
\label{Fig_Result2}
\end{figure*}

In this section, we clarify how to select the hyper-parameters of a deep recurrent network to obtain high prediction accuracy and then make use of Monte-Carlo simulations to evaluate the outage probability and channel capacity of PRS, compared with the existing schemes including ORS and OSTC.  Following the channel assumption adopted by most of the previous works in this field,  we would apply single-antenna  flat-fading  \textit{i.i.d.} channels. Each channel follows the Rayleigh distribution with an average power gain of $0\mathrm{dB}$, where its fading coefficient $h {\sim} \mathcal{CN}(0, 1)$. The default maximal Doppler frequency shift  is set to $f_d{=}100 \mathrm{Hz}$, emulating  fast fading environment. Continuous-time channel responses are sampled with a rate of $f_s{=}1 \mathrm{KHz}$, adhering to the assumption of flat fading, and therefore the interval of samples is  $T_s{=}1\mathrm{ms}$. Each channel generates a series of $10^6$ consecutive samples $ \{h[t]\left |  t{=}1,2,\ldots,10^6 \right.  \}$. As usual, an $\mathbb{EE}$ target rate of $R{=}1 \mathrm{bps/Hz}$ is applied for outage calculation. The total transmit power $P$ is equally allocated between two phases, where the source's power is $P_s{=}0.5P$, resulting in an  average SNR $\bar{\gamma}_{s,k}{=}0.5P/\sigma_n^2$, while $\bar{\gamma}_{k,d}{=}0.5P/\sigma_n^2$ for the $\mathbb{RD}$ link.

\subsection{Training the Predictor}
The hyper-parameters of a deep network such as the number of layers or neurons have a substantial impact on prediction accuracy. It is worth clarifying how to tune a deep network on demand. A training process starts from an initial state where all weights and biases are randomly selected.  The input of the predictor at the relay is $a_{k,d}[t]$ and the output is its $D$-step-ahead prediction $\check{a}_{k,d}[t{+}D]$. To measure prediction accuracy, the mean squared error (MSE)  is applied as the cost function, namely $
\mathrm{MSE} = \frac{1}{T} \sum_{t=1}^ T \left |  a_{k,d}[t{+}D] - \check{a}_{k,d}[t{+}D] \right |^2$, where $T$ is the total number of channel samples for evaluation.  Using the \emph{batch} training, a batch of $256$ samples is fed into the network per step. The output is compared with the desired values and the resultant error signals are propagated back through the network to update the weights by means of training algorithms such as the Adam optimizer used in our simulation. After $10$ epochs, the trained network is employed to predict CSI.

\figurename \ref{Fig_MSE} compares the prediction accuracy of the predictors with different hyper-parameters.  Let's first look at the impact of the number of layers and the number of neurons. Starting from an LSTM network with  a single hidden layer, denoted by \emph{LSTM-1} in the legend of the figure, its accuracy curve  as a function of the number of hidden neurons likes an `U' shape. That is because the network suffers from the \emph{under-fitting} problem with only $20$ neurons in the hidden layer, while the \emph{over-fitting} problem appears using over $80$ neurons.  To make a fair comparison, the horizontal axis represents the total number of hidden neurons, which are evenly allocated across layers. For instance, the point of `60' in the horizontal axis means a 2-hidden-layer network with $30$ neurons at either layer (denoted by \emph{LSTM-2}), a 3-hidden-layer network with $20$ neurons per layer (denoted by \emph{LSTM-3}),  or  a single layer with $60$ hidden neurons.  No matter how many neurons in its single hidden layer, \emph{LSTM-1} cannot reach the high accuracy achieved by \emph{LSTM-2} and \emph{LSTM-3},  justifying the benefit of deep learning. But it does not mean that the more layers, the better,  as shown by the worse result of \emph{LSTM-4}, which has 4 hidden layers. After known that 2-hidden-layer is the best choice for LSTM, we further observe the recurrent networks with 2 RNN or GRU hidden layers, indicated by \emph{RNN-2} and \emph{GRU-2}, respectively. As we can see, GRU performs as good as LSTM, whereas RNN is weak. As a result, we select a 2-hidden-layer LSTM network with $25$ neurons at either layer, upon which the numerical results in the following figures are derived.
\subsection{Performance Comparison}
We further compare the outage performance of three relaying schemes in a cooperative network with $K{=}8$ relays, as illustrated in \figurename \ref{Fig_DF_outage}. The relay selection with the perfect knowledge of CSI (i.e., $\rho{=}1$) is used as the benchmark, which has the diversity order of $8$ and decays at a rate of $1/\bar{\gamma}^8$, where $\bar{\gamma}=P/\sigma_n^2$ is the average $\mathbb{EE}$ SNR. With the delay of $\tau=2$ and $3\mathrm{ms}$, the quality of outdated CSI drops to $\rho_o=J_0(0.4\pi)\thickapprox 0.6425$ and $J_0(0.6\pi)\thickapprox0.2906$, respectively, which substantially deteriorates the performance. The diversity of ORS falls into $1$, i.e., no diversity, and the curve decays slowly  at a rate of $1/\bar{\gamma}$ in the high SNR regime.    OSTC can redeem some loss and achieve the diversity order of $2$ by using a pair of relays,  but its gap to the benchmark is still large, more than $7\mathrm{dB}$ at the level of $10^{-3}$. Making use of channel prediction, the quality of CSI can be improved to $\rho>0.95$. The proposed scheme achieves nearly the optimal performance with the horizon of $2\mathrm{ms}$ (by setting $D=2$ steps prediction), and remarkably outperforms OSTC with a gain of approximately $8\mathrm{dB}$ in the case of $3\mathrm{ms}$.
Moreover, the channel capacities for different schemes  given $\tau=3\mathrm{ms}$ are comparatively illustrated in \figurename \ref{Fig_capacity}. At the SNR of $\bar{\gamma}{=}20\mathrm{dB}$, for instance, ORS, OSTC, and PRS achieves $2.6$, $2.75$, and $3.5\mathrm{bps/Hz}$, respectively, where ORS suffers from a loss of around $1\mathrm{bps/Hz}$ but PRS achieves a near-optimal capacity.

\section{Conclusions}
In this paper, we proposed a deep-learning-aided cooperative diversity method for mobile terminals without an antenna array to cultivate the benefit of spatial diversity. A recurrent neural network was deliberately built to improve the timeliness of channel state information applied for selecting a single opportunistic relay. Simply inserting a channel predictor between the channel estimator and relay selector, an ORS system can be upgraded to a PRS system without any other modifications, making it transparent and easier to compatible with the existing systems and standards. It achieves the optimal performance with the full diversity order equaling to the number of cooperating relays in slow fading wireless environments, and  substantially outperforms the existing schemes in fast fading channels.  It inherits the simplicity of ORS by avoiding multi-relay coordination and synchronization, and the computational complexity arising from fading channel prediction is negligible compared with COTS hardware. From the perspective of \textit{performance},   \textit{compatibility}, and \textit{ complexity}, it is viewed as a good candidate for next-generation cooperative networks.

\bibliographystyle{IEEEtran}
\bibliography{IEEEabrv,Ref_WCNC2021}
\end{document}